\documentclass{elsarticle}
\usepackage{graphicx}
\usepackage{amsmath,amsfonts,amssymb}

\journal{Applied Mathematical Modelling}

\newcommand{\derp}[2]{\frac{\partial #1}{\partial #2}}
\newcommand{\der}[2]{\frac{d #1}{d #2}}

\def \diver{\nabla \cdot}

\begin{document}

\begin{frontmatter}
\title{Modeling water uptake by a root system growing in a fixed soil 
volume.}

\author{J. L. Blengino Albrieu}
\address{Departamento de F\'{\i}sica, Facultad de Ciencias Exactas 
F\'{\i}sico-Qu\'{\i}micas y Naturales, Universidad Nacional de  R\'{\i}o 
Cuarto, Ruta 8 km 601, X5804BYA R\'{\i}o Cuarto, C\'ordoba, ARGENTINA}
\ead{jblengino@exa.unrc.edu.ar}

\author{J. C. Reginato}
\address{Departamento de F\'{\i}sica, Facultad de Ciencias Exactas
F\'{\i}sico-Qu\'{\i}micas y Naturales, Universidad Nacional de
R\'{\i}o Cuarto, Ruta 8 km 601, X5804BYA R\'{\i}o Cuarto,
C\'ordoba, ARGENTINA}

\author{D. A. Tarzia}
\address{Departamento de Matem\'atica - CONICET, Facultad de Ciencias
Empresariales, Universidad Austral, Paraguay 1950, S2000FZF
Rosario, ARGENTINA}

\begin{abstract}
The water uptake by roots of plants is examined for an ideal situation, 
with an approximation that resembles plants growing in pots, meaning that 
the total soil volume is fixed. We propose a coupled water uptake-root 
growth model. A one-dimensional model for water flux and water uptake by a 
root system growing uniformly distributed in the soil is presented, and 
the Van Genuchten model for the transport of water in soil is used. The 
governing equations are represented by a moving boundary model for which 
the root length, as a function of time, is prescribed. The solution of the 
model is obtained by front-fixing and finite element methods. Model 
predictions for water uptake by a same plant growing in loam, silt and 
clay soils are obtained and compared. A sensitivity analysis to determine 
relative effects on water uptake when system parameters are changed is 
also presented and shows that the model and numerical method proposed are 
more sensitive to the root growth rate than to the rest of the parameters. 
This sensitivity decreases along time, reaching its maximum at thirty 
days. A comparison of this model with a fixed boundary model with and 
without root growth is also made. The results show qualitative differences 
from the beginning of the simulations, and quantitative differences after 
ten days of simulations.
\end{abstract}


\end{frontmatter}

\section{Introduction}

In the development of a theory to describe plant water uptake, electrical 
analogues  of the system have been used for analysis 
\cite{cowan1965,gardner1960,vandenhonert1948}. The analogues are based on 
the assumption that rooting patterns are uniform and constant in each soil 
layer. Steady flow is presumed in both the soil and the plant over the 
period of calculation. In this approach the plant water potentials are 
primarily the result of an imposed value of transpiration rate and its 
variations. Later papers \cite{feddes2001,hillel1980,molz1981} have
presented detailed reviews on plant water uptake. In those papers the 
Richards equation is used, with a sink term. Another approach is to model 
the water movement and uptake over large areas, using individual plant 
\cite{guswa2005}, or global behavior \cite{puma2005}. A microscopical 
approach has also been proposed \cite{personne2003}, where the total water 
uptake is calculated based on using a constant value for the entire 
rooting profile. In more recent papers \cite{roose2004a,wang2004,wu1999} 
the root growth has been taken into account, still using a fixed domain. 
The root growth is inscribed on a domain that is not a function of time. 
Some other papers about nutrient uptake consider root growth and 
instantaneous coupling with the nutrient flux by using a variable domain 
approximation \cite{reginato2000}. In this last model \cite{reginato2000} 
a variable root length, and consequently, a variable available volume of 
soil to each root of a root system is considered using a moving boundary 
model. In this model the root system is uniformly distributed in the soil 
and the variation of available soil volume per unit of root length is 
modeled by a moving boundary. The approach presented here is based on that 
in \cite{reginato2000} . In the proposed model plants growing in 
controlled conditions, as in a growth chamber, are assumed. A constant 
temperature and evapotranspiration rate is presumed. In this situation, 
the water potential at the root surface is determined by the soil water 
potential, and consequently determines water uptake by the growing root 
system. The proposed model considers an uniform root water uptake for all 
the root system. The goal of this paper is to present a simplified model 
of water uptake coupled with a growing root system and analyze the 
influence of system parameters on water uptake using typical values.

\section{Model}

Darcy's law describes the flow of water on a porous unsaturated medium as
\begin{equation}\label{eq:flujo1}
  \vec J(\vec r, t) = -K(\Psi(\vec r, t)) \vec \nabla \Psi(\vec r, t)
\end{equation}
with $\vec J[cm^3/cm^2 s]$ the water flux per surface unit at position 
$\vec r [cm]$ at time $t[s]$, $K [cm/s]$ the soil water conductivity and 
$\Psi [cm]$ the soil water potential.

The corresponding continuity equation (mass conservation) is given by
\begin{equation}
  -\diver \vec{J} = \derp\theta t,
\end{equation}
with $\theta [cm^3/cm^3]$ the soil water content per unit of volume, with 
the approximation of only radial flux the transport equation results
\begin{equation}\label{eq:cont1}
  \derp{}t\left[\Psi(r,t)\right]=-\frac1{rC(\Psi(r,t)) } \derp{}{r} 
  \left[r K(\Psi(r,t)) \derp{}r \left[\Psi(r,t)\right]\right]
\end{equation}
where $r$ is the cylindrical radial coordinate, and
\begin{equation}
  C(\Psi(r,t))=\der\theta\Psi(\Psi(r,t)) 
\end{equation}
is the differential capacity of water $[cm^{-1}]$.

The soil water constitutive relations for $K$, $\theta$, and $C$, as 
functions of $\Psi$, are the ones proposed by Van Genuchten 
\cite{vangenuchten1980}, and consist of the expressions given by:
\begin{eqnarray}
  K(\Psi)&=&K_s \frac{\left\{\left[1+\left(\frac\Psi{\Psi_e} \right)^n
  \right]^m - \left(\frac\Psi{\Psi_e} \right)^{n-1}\right\}^2}
  {\left[1+\left(\frac\Psi{\Psi_e}\right)^n\right]^{m(p+2)}},
  \label{eq:kapsivG}\\
  \theta(\Psi) &=& \left[1+\left(\frac{\Psi}{\Psi_e}\right)^n\right]^{-m} 
  (\theta_s-\theta_R) + \theta_R, 
  \label{eq:titapsivG}\\
  C(\Psi) &=& (\theta_s-\theta_R) \frac{1-n}{\Psi_e} \left[1+\left(
  \frac{\Psi}{\Psi_e}\right)^n\right]^{-m-1} \left(\frac{\Psi}
  {\Psi_e}\right)^{n-1}, 
  \label{eq:CpsivG}
\end{eqnarray}
where $K_s [cm/s]$ is the saturated soil conductivity, $\theta_s 
[cm^3/cm^3]$ is the saturated soil water content, $\theta_R [cm^3/cm^3]$ 
is the residual water content, $\Psi_e [cm]$, $p [1]$ and $n [1]$ are 
experimental coefficients, and $m = 1-1/n$.

\begin{figure}[tp]
  \center{\noindent\includegraphics[width=11cm]{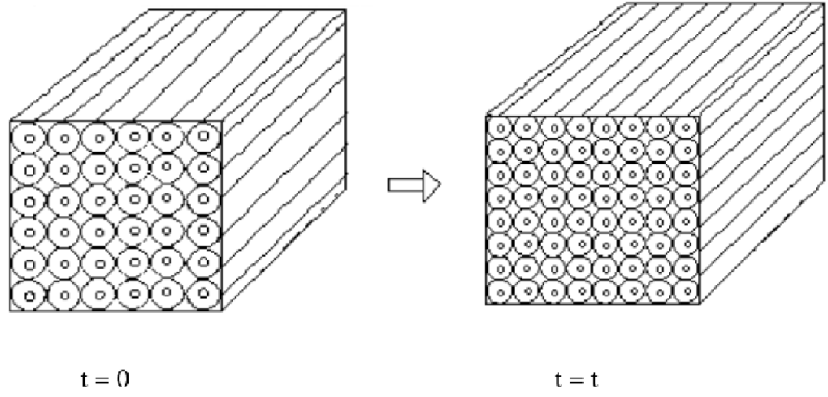}}
  \caption{Homogeneous rooting in soil and its time evolution.}
  \label{fig:evolucion}
\end{figure}

\begin{figure}[tp]
  \center{\noindent\includegraphics[width=90mm]{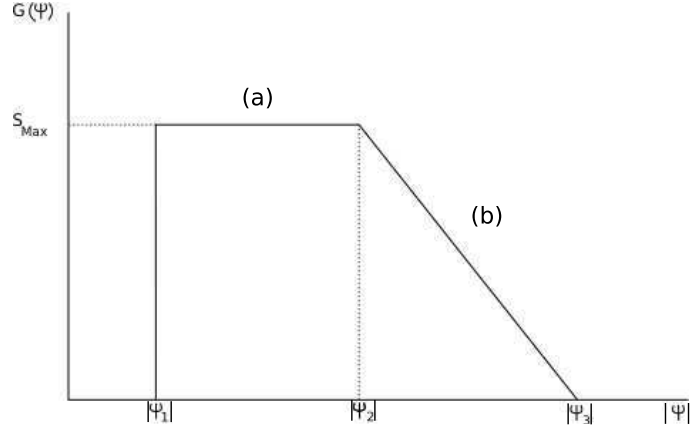}}
  \caption{Root water uptake function as proposed by Feddes 
  \citep{feddes2001}. The zone (a) is the zone of maximum water uptake, 
  while the zone (b) is the zone of linear water uptake.}
  \label{fig:tomafig}
\end{figure}

\begin{figure}[tp]
  \center{\noindent \includegraphics[width=60mm]{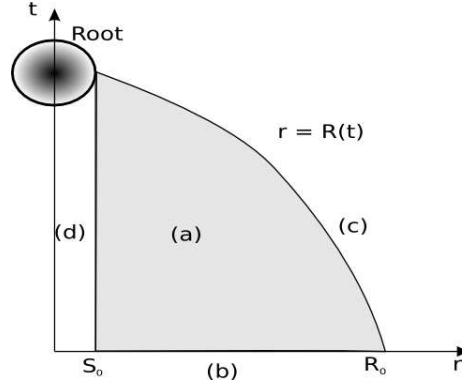}}
  \caption{model of the domain of validity of the proposed model, the 
  shaded zone (a) represents the zone where the equation 
  (\ref{eq:modelo1}) is valid. The line (b) is the initial value condition 
  (\ref{eq:modelo2}). In the curve (c) $r=R(t)$ the boundary condition 
  (\ref{eq:modelo3}) is used. In the line (d) the root water uptake 
  condition (\ref{eq:modelo4}) is used.}\label{fig:dominio}
\end{figure}

We presume that the water potential does not change with soil depth, the 
soil does not evaporate, and laboratory conditions, like light and 
temperature are maintained constant. The root density is homogeneous on 
the soil, the total volume is fixed (as in pots), therefore the soil 
volume per unit of root length is decreasing as in Figure 
\ref{fig:evolucion}. Based on these assumptions, and taking into account 
the root length density as a function of time $(t)$ the following moving 
boundary model in cylindrical coordinates \cite{personne2003,reginato2000} 
is proposed
\begin{eqnarray}
  \derp{}t\left[\Psi(r,t)\right]&=&-\frac1{rC(\Psi(r,t)) }\derp{}{r} 
  \left[r K(\Psi(r,t)) \derp{}r \left[\Psi(r,t)\right]\right]
  \label{eq:modelo1}\\
  \Psi(r,0) &=& \phi(r)
  \label{eq:modelo2}\\
  H\left(\Psi(R(t),t),t\right)&=&2\pi R(t)K(\Psi(R(t),t))\derp{\Psi}{r}
  (R(t),t)
\label{eq:modelo3}\\
  G\left(\Psi(s_0, t)\right)&=&-2\pi s_0 K(\Psi(s_0,t))\derp{\Psi}{r}
  (s_0,t)
\label{eq:modelo4}\\
  R(t)&=&R_0\sqrt{\frac{l_0}{l(t)}}\label{eq:modelo5}
\end{eqnarray}
with $s_0\leq r\leq R(t)$ and $0\leq t\leq T_{max}$, where $T_{max}$ is 
the maximum time for which the system has meaning $\left(R(T_{max})\geq 
s_0\right)$, $s_0$ is the root radius, $R_0$ is the initial half-distance 
among roots, $R(t)$ is the instantaneous half-distance among roots (a  
decreasing function as root density grows), $l_0$ is the initial root 
length, and $l(t)$ is the instantaneous root length. Equation 
(\ref{eq:modelo1}) is the pressure head based Buckingham-Richards 
equation. The condition (\ref{eq:modelo2}) is the initial water potential 
profile, with $\phi(r)$ a single valued function. The condition 
(\ref{eq:modelo3}) represents the flux $(H)$ on the moving boundary $r = 
R(t)$, which will be considered null in this paper as an approximation to 
a soil isolated. The condition (\ref{eq:modelo4}) is the boundary 
condition at the root soil interface $(r = s_0)$ representing the root 
water uptake per unit of root length $\left( G(\Psi(s_0,t))[cm^3/cm\ 
s]\right)$. For the water uptake function $\left( G(\Psi)\right)$ we use 
the function proposed by Feddes \cite{feddes2001} which is given by:
\begin{center}
  \begin{tabular}[b]{lllcl}
    \\
    $G(\Psi) = 0$ &if &$0>\Psi \geq \Psi_1$ \\
    $G(\Psi) = S_{max}$ &if &$\Psi_1>\Psi\geq\Psi_2$\\
    $G(\Psi) = S_{max}\frac{\Psi -\Psi3}{\Psi_2 -\Psi_3}$ &if&
    $\Psi_2>\Psi\geq\Psi_3$\\
    $G(\Psi) = 0$ &if &$\Psi_3>\Psi$ \\
  \end{tabular}
\end{center}
where $\Psi_1$, $\Psi_2$ and $\Psi_3$ are the anaerobiosis point, the 
limiting point, and the wilting point, respectively. $S_{max} \left[cm^2/s
\right]$ is the maximum water uptake per unit of root length. A diagram of 
this function can be seen on Figure \ref{fig:tomafig}. The condition 
(\ref{eq:modelo5}) is the time dependence of the moving boundary $(R(t))$ 
which is obtained presuming a fixed total volume including soil and roots, 
and a linear growth rate $(l(t) = l_0 + V \ t)$ \cite{reginato2000}, where 
$V \left[cm/s\right]$ is the root length growth rate. Graphical evolution 
of this system with time can be seen in Figure \ref{fig:evolucion}. A 
schematic mathematical diagram of the problem is shown in Figure 
\ref{fig:dominio}. Once the water potential $\Psi(s_0,t)$ at the root 
surface is obtained, the water uptake is computed with a variable domain 
integration method given by \cite{reginato2002}
\begin{eqnarray}
	U(t)&=&l_0\int_0^{t}G(\Psi(s_0,\tau))d\tau\nonumber\\&&+\int_0^{t}
	\left(\int_\tau^{t}G(\Psi(s_0,\tau))d\tau\right)\dot l(\tau)d\tau
\end{eqnarray}
where $U \left[cm^3\right]$ is the cumulative water uptake at time $t$. 
This last expression can be simplified to (see \ref{app1})
\begin{eqnarray}\label{eq:tarziasimp}
  U(t) &=& \int_0^t G(\Psi(s_0,\tau)) l(\tau) d\tau,
\end{eqnarray}
then the instantaneous water uptake can be defined as
\begin{equation}
 U_i(t)=\dot{U}(t) = G(\Psi(s_0,t)) l(t).
\end{equation}
The remaining soil water $\left( W \left[cm^3\right]\right)$ as a function 
of time can be calculated using
\begin{equation}
  W(t) = 2 \pi\ l(t)\int_{s_0}^{R(t)}\theta\left(\Psi(r,t)\right) r\ dr,
  \label{eq:aguasuelo}
\end{equation}
if the root is not growing then $l(t) = l_0 = constant$, and as a 
consequence $R(t) = R_0$, therefore the water remaining in the soil in 
this case is
\begin{equation}
  W(t)=2\pi\ l_0\int_{s_0}^{R_0}\theta\left(\Psi(r,t)\right)r\ dr.
  \label{eq:aguasuelofija}
\end{equation}

To  solve the system (\ref{eq:modelo1})--(\ref{eq:modelo5}) the domain is 
transformed to a dimensionless form by using the following expressions:
\begin{eqnarray}
  x &=& \frac{r-s_0}{R(t)-s_0}, \nonumber \\
  \tau &=& \frac t{t_0},\label{eq:cambiovariables}\\ \nonumber
  \Phi(x,\tau)&=&\frac{\Psi(r,t)}{\Psi_e}=\frac{\Psi(x(R(t)-s_0)+s_0,\tau 
  t_0)}{\Psi_e},
\end{eqnarray}
or their inverses given by
\begin{eqnarray}
  r &=& s_0 +(R(t) - s_0) x \nonumber \\
  t &=& t_0 \tau \\ \nonumber
  \Psi(r,t)&=&\Psi_e\Phi\left(x,\tau\right)=\Psi_e\Phi\left(\frac{r-s_0}
  {R(t)-s_0},\frac t{t_0}\right).
\end{eqnarray}
and the following definitions
\begin{eqnarray}
  k(\Phi(x,\tau)) &=& \frac{K(\Psi(r,t))}{K_s} = \frac{\left\{\left[1+
  \Phi^n(x,\tau)\right]^m - \Phi^{n-1}(x,\tau)\right\}^2}{\left[1+\Phi^n
  (x,\tau)\right]^{m(p+2)}}
  \label{eq:kadim}\\
  c(\Phi(x,\tau))&=&  \frac{C(\Psi(r,t))\Psi_e}{(\theta_s-\theta_R) (1-n)} 
  =\frac{\Phi^{n-1}(x,\tau)}{\left[1+\Phi^n(x,\tau)\right]^{m+1}}
  \label{eq:cadim}\\
  v(\tau) &=& \frac{R(t)}{s_0}
  \\
  w(\tau) &=& \frac{l(t)}{s_0}
  \\
  g(\Phi(x,\tau))&=&\frac{G(\Psi(r,t))}{S_{max}}
  \\
  H(\Psi(r,t))&=&0,
  \\
  c^*(\Phi(x,\tau),x,\tau)&=&\left((v(\tau)-1)x+1\right) c(\Phi(x,\tau))
  \\
  k^*(\Phi(x,\tau),x,\tau)&=&\left((v(\tau)-1)x+1\right) k(\Phi(x,\tau))
  \\
  \tilde c(\Phi(x,\tau),x,\tau)&=&-\frac{xv(\tau)w^\prime(\tau)}{2w(\tau)
  (v(\tau)-1)}c^*(\Phi(x,\tau),x,\tau)
  \\
  v(\tau)&=&\frac{R_0}{s_0}\sqrt{\frac{l_0}{s_0 w(\tau)}}\label{eq:vadim}
  \\
  \gamma&=&\frac{t_0\Psi_e K_s}{(\theta_s-\theta_R)(1-n)s_0^2}
  \label{eq:gamma}
  \\
  \sigma &=& \frac{S_{max}}{2 \pi K_s \Psi_e}\label{eq:sigma}
  \\
  \tau_{max} &=& \frac{T_{max}}{t_0}\label{eq:tmax}
\end{eqnarray}
where $k(\Phi(x,\tau))$ and $c(\Phi(x,\tau))$ are the dimensionless forms 
of $K(\Psi(r,t))$ and $C(\Psi(r,t))$ respectively. The dimensionless 
number $\gamma$ takes into account soil properties and the geometrical 
proportions of the root, $t_0$ is chosen to make $\gamma = 1$. The 
dimensionless number $\sigma$ takes into account one flux per unit of 
length, related to the water uptake, the term $K_s \Psi_e$ is a flux per 
unit of length when the soil is saturated and the gradient of the matric 
potential is equal to $\Psi_e/1cm$.

Taking into account (\ref{eq:cambiovariables}) and the definitions 
(\ref{eq:kadim})--(\ref{eq:tmax}), the system 
(\ref{eq:modelo1})--(\ref{eq:modelo5}) is transformed in a dimensionless 
form in the domain $(0,1) \times (0,\tau_{max})$ given by:
\begin{eqnarray}
\label{eq:sistema-cuerpo}
  c^*(\Phi(x,\tau),x,\tau)\derp{}{\tau}\left[\Phi(x,\tau)\right]&=&\tilde 
  c(\Phi(x,\tau),x,\tau) \derp{}{x}\left[\Phi(x,\tau)\right]-\\&&\nonumber
  \derp{}{x}\left[\frac{\gamma k^*(\Phi(x,\tau),x,\tau)}{(v(\tau)-1)^2}
  \derp{}{x}\left[\Phi(x,\tau)\right]\right],
  \\
  \label{eq:sistema-cond-in}
  \Phi(x,0) &=& \varphi(x),
  \\
  \label{eq:sistema-borde-lej}
  \derp{\Phi}{x} (1,\tau) &=& 0,
  \\
  \label{eq:sistema-absorcion}
  \sigma\  g(\Phi(0,\tau))&=&-\frac{1}{v(\tau)-1} k^*(\Phi(0,\tau),0,\tau)
  \derp{\Phi}{x}(0,\tau),
\end{eqnarray}
where $\varphi(x) = \phi(s_0+(R_0-s_0)x)/\Psi_e$, is the dimensionless 
initial profile. This transformation maps the spatial domain variable in 
time to a fixed domain in time, and adds a term on the right side of the 
transformed transport equation (\ref{eq:sistema-cuerpo}) which contains 
the variation of the moving boundary. In order to solve the model 
(\ref{eq:sistema-cuerpo}--\ref{eq:sistema-absorcion}) the non-linear 
finite element method \cite{reddy2004} is applied and the resulting model 
is solved by using the software FlexPDE \cite{flexpde} with an adaptive 
mesh of around 400 nodes. 

\section{Results}

All simulations were performed  by a same hypothetical plant in three 
types of soils (loam, silt and clay) for the same soil-root volume $(4000 
cm^3)$ with the same total water content $(\theta_i=0.30)$. This initial 
condition is fixed using the "field capacity" concept given by Ritchie 
\cite{ritchie1981}. Hydraulic soil data selected were those for loam, 
silt, and clay based on \cite{personne2003}. The soil parameters used are 
shown in Table \ref{tab:parametros_suelo}.  For the plant parameters 
different sources were used. The values of $\Psi_2$ and $\Psi_3$ were 
taken from \cite{feddes1978}, the value of $\Psi_1$ was chosen to assure 
that wateruptake was possible at the beginning of the simulation. $s_0$ 
and $V$ are typical values from the literature \cite{kelly1992}, and $l_0$ 
was chosen to be $1$ cm to simulate a plant at the start of growth. The 
plant and soil volume parameters are listed in Table \ref{tab:tabla1}. 
Table \ref{tab:parametros_iniciales} shows parameters of soil and plant 
properties and the initial ``available water''$(U_0)$. This last parameter 
is approximated as
\begin{equation}
  U_0 = [\theta_i -\theta(\Psi_3)] \times 4000 cm^3,
\end{equation}
and represents how much water can be extracted from the soil before it 
reaches an uniform water potential at the wilting point. Beyond this point 
the root cannot extract more water.

\begin{table}
  \begin{flushleft}
    \begin{tabular}{l|c|c|c|c|c|c}
      Soil&$\theta_s$&$\theta_R$&$K_s$ (cm/s)& $\Psi_e$ (cm)&$n$& $p$\\
      \hline
      Loam&$0.43$&$0.078$&$2.89\times 10^{-4}$&$-27.78$&$1.56$&$0.5$\\
      Silt&$0.46$&$0.034$&$7.00\times 10^{-5}$&$-62.5$&$1.37$&$0.5$\\
      Clay&$0.38$&$0.068$&$5.80\times 10^{-6}$&$-125$&$1.09$&$0.5$
    \end{tabular}
  \end{flushleft}
  \caption{Soil hydraulic properties}\label{tab:parametros_suelo}
\end{table}

\begin{table}
  \begin{flushleft}
    \begin{tabular}{c|ll}
      Parameter&Value\\ \hline
      $l_0$&1&cm\\
      $R_0$&35.7&cm\\
      $V$&$1 * 10^{-4}$&cm/s\\
      $s_0$&0.05&cm\\
      $S_{max}$&$2*10^{-6}$ &cm$^2$/s \\
      $\Psi_1$&-1&cm\\
      $\Psi_2$&-750 &cm\\
      $\Psi_3$&-17500&cm
    \end{tabular}
  \end{flushleft}
  \caption{Values for the plant and soil volume parameters}
  \label{tab:tabla1}
\end{table}

\begin{table}
  \begin{flushleft}
    \begin{tabular}{l|c|c|c}
      Soil & $\Psi(r,0)$ (cm) & $t_0$ (s) & $U_0$ (cm$^3$)\\
      \hline
      Loam & $-51.4$ &$0.0614$ & $850$ \\
      Silt & $-194$ & $0.0901$ & $852$\\
      Clay & $-3280$ &$0.0968$ & $128$
    \end{tabular}
  \end{flushleft}
  \caption{Simulation parameters depending on soil and plant properties}
  \label{tab:parametros_iniciales}
\end{table}

\begin{figure}[tp]
  \center{\noindent\includegraphics[scale=0.3]{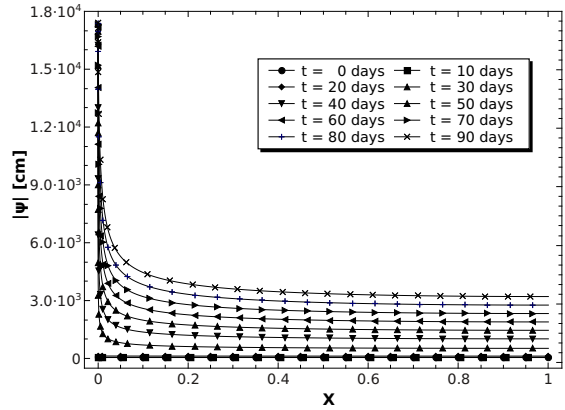}}
  \caption{Water potential profiles at the scaled domain for the loam 
  soil.}\label{fig:loam_profile}
\end{figure}

\begin{figure}[tp]
  \center{\noindent\includegraphics[scale=0.3]{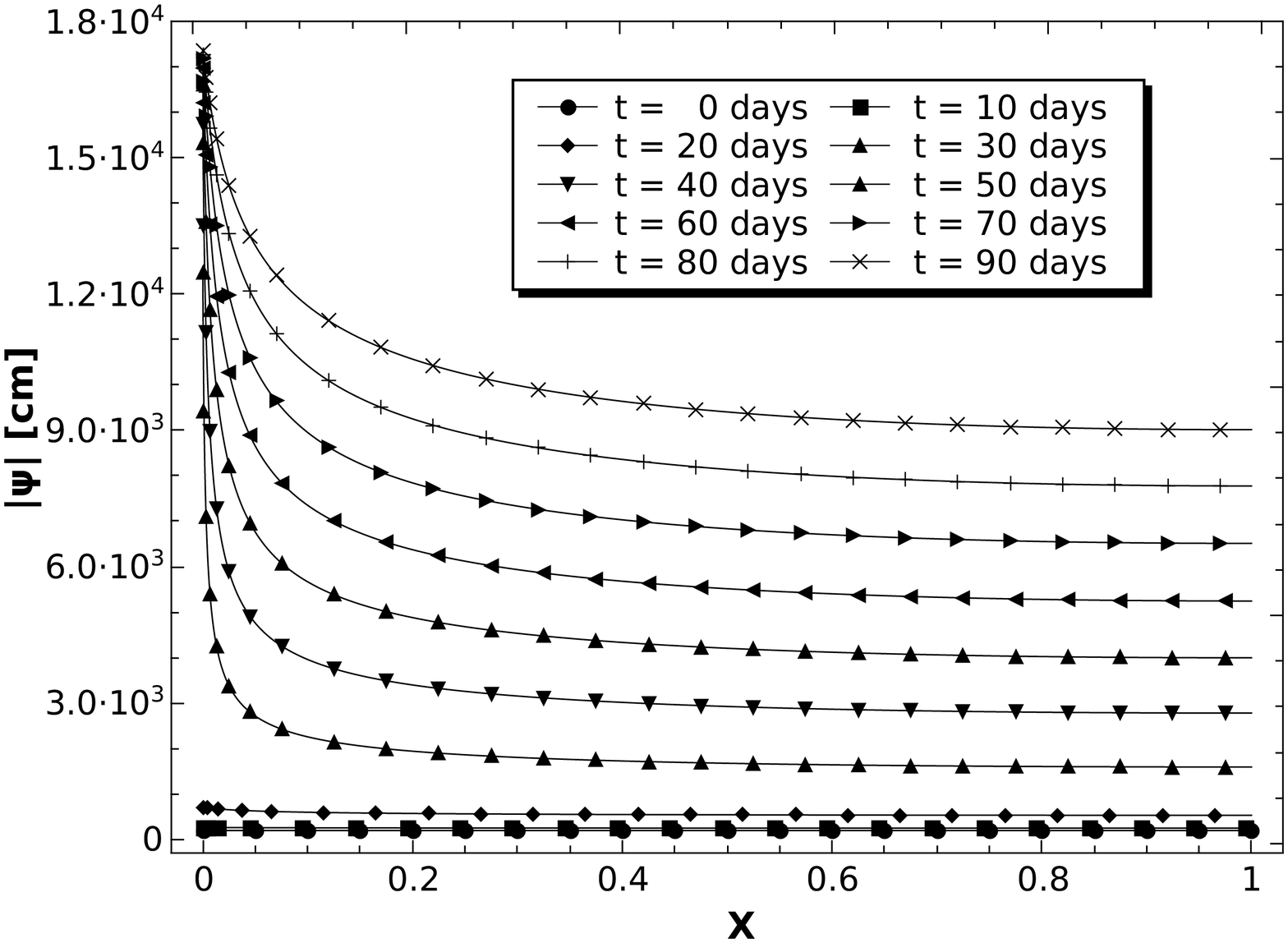}}
  \caption{Water potential profiles at the scaled domain for the silt 
  soil.} \label{fig:silt_profile}
\end{figure}

\begin{figure}[tp]
  \center{\noindent\includegraphics[scale=0.3]{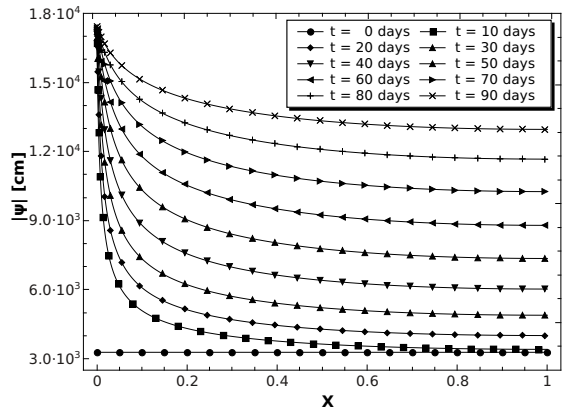}}
  \caption{Water potential profiles at the scaled domain for the clay 
  soil.} \label{fig:clay_profile}
\end{figure}

Figures \ref{fig:loam_profile}, \ref{fig:silt_profile}, and 
\ref{fig:clay_profile} show the soil water potential profiles at different 
times, with an initial moisture condition equivalent $(\theta=0.30)$, for 
the loam, silt and clay soil, respectively. The curves reveal that a high 
water potential gradient is developed in a very small time period for the 
clay soil, while for the loam and silt soil the development of the water 
potential gradient is more gradual. For loam and silt soil the root dries 
the soil near the root surface on the first 30 days. After that period 
these soils show a water potential on the root surface close to the 
wilting point. The root starts to retrieve water near the moving boundary 
$R(t)$. This is shown by an increase of the modulus of the water potential 
(decrease of his value) at $x=1$ for the dimensionless domain, or $r = 
R(t)$ for the physical domain. The clay soil shows a similar behavior but 
the time period at which the potential near root zone reaches potentials 
close to the wilting point is 10 days.

\begin{figure}[p]
  \center{\noindent\includegraphics[width=8cm]{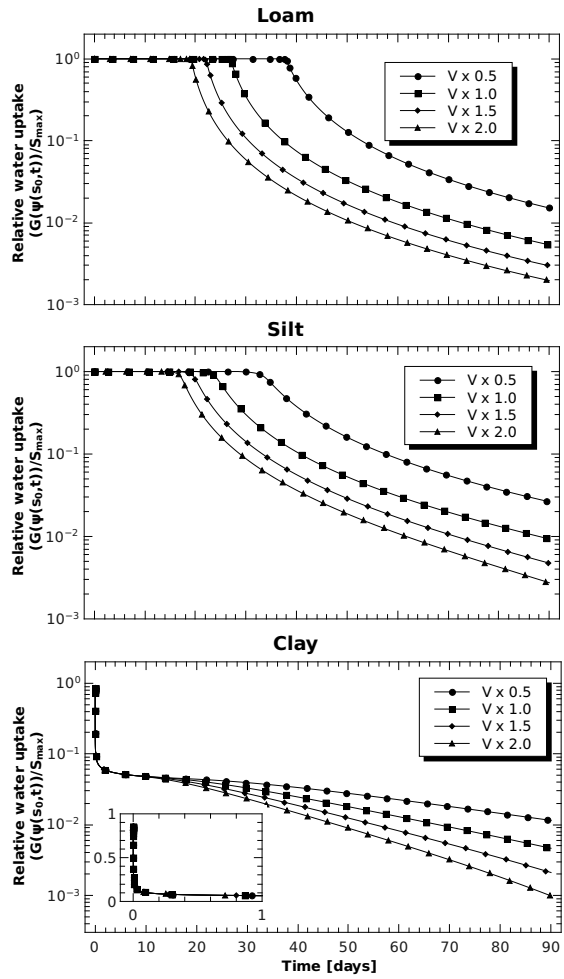}}
  \caption{Relative water uptake per unit of root length time evolution 
  for the different soils. The curves refer to changes in root growth 
  speed $(V)$. The square is a zoom on the first day of simulation for the 
  clay soil.}\label{fig:sens_g_v}
\end{figure}

Figure \ref{fig:sens_g_v} shows the relative water uptake 
$(G(\Psi(s_0,t))/S_{max})$ per unit of root length as a function of time 
for the three types of soil, at different root growth rate. At the initial 
time the water uptake for loam and silt soil develops at the zone of 
maximum uptake (zone (a) in Figure \ref{fig:tomafig}), while clay soil 
does the same for the linear uptake zone (zone (b) in Figure 
\ref{fig:tomafig}). The figure shows that the water uptake does not vary 
much for the first 20 days as a function of $V$ for loam soil, after that 
period loam soil enters on the linear regime at different times depending 
on the value of $V$. After 40 days the gap among the curves on the log 
scale remains almost constant and all the curves are in the linear uptake 
regime. This means that the instantaneous root water uptake curves have a 
multiplicative constant among them. Silt soil has a very similar behavior 
to loam soil but the linear regime occurs at shorter times. For clay soil 
the period of similar water uptake is 10 days and after that period the 
curves begin a gradual separation that continues until the end of the 
simulation. Clay soil also shows a very sharp initial decrease on the 
curve (one order of magnitude), this is due to the linear regime and a 
large decrease of $\Psi(s_0,t)$ at the beginning of the simulation.

\begin{figure}[p]
  \begin{center}
    \begin{tabular}{c}
      \includegraphics[scale=0.29]{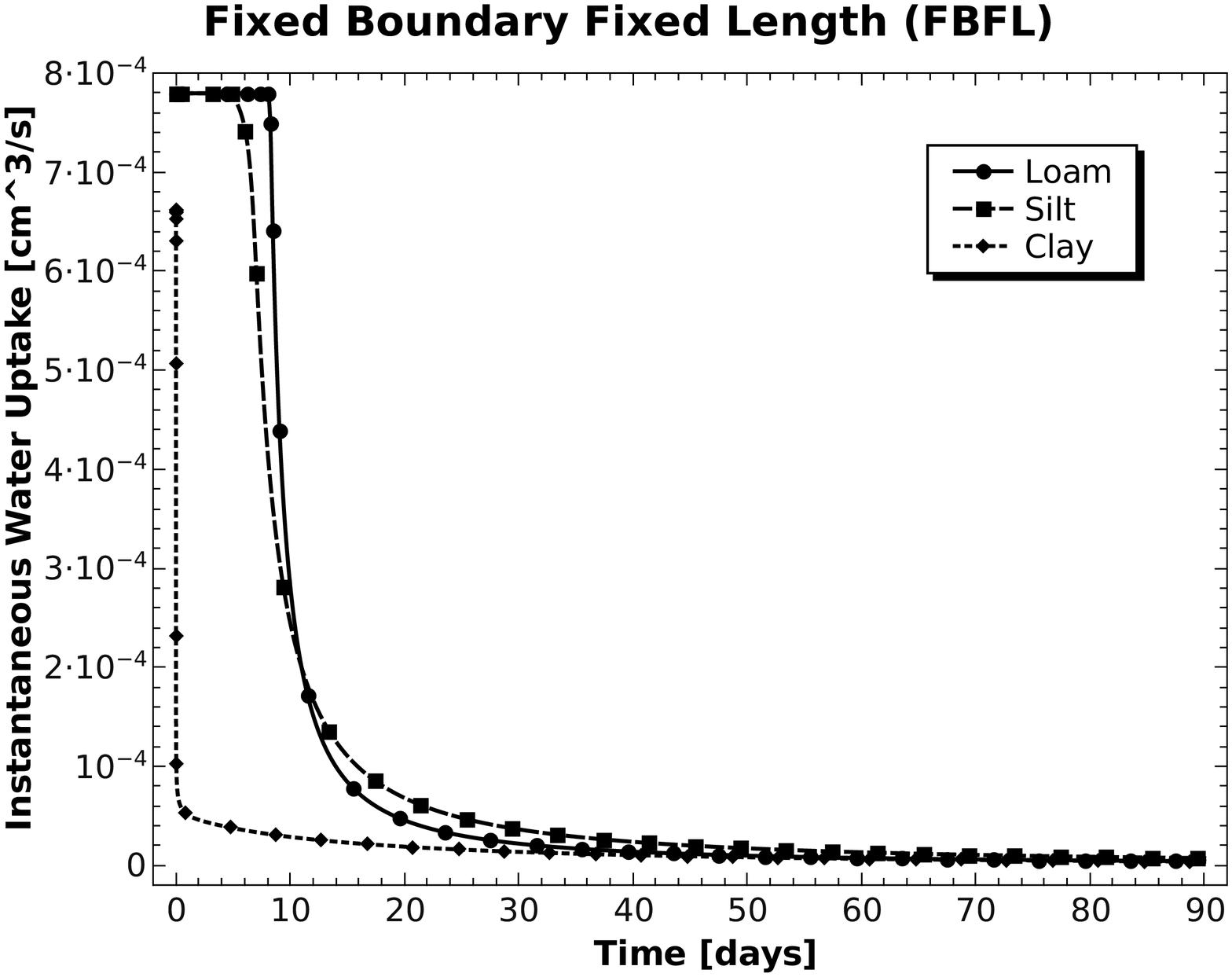}
      \\
      \includegraphics[scale=0.29]{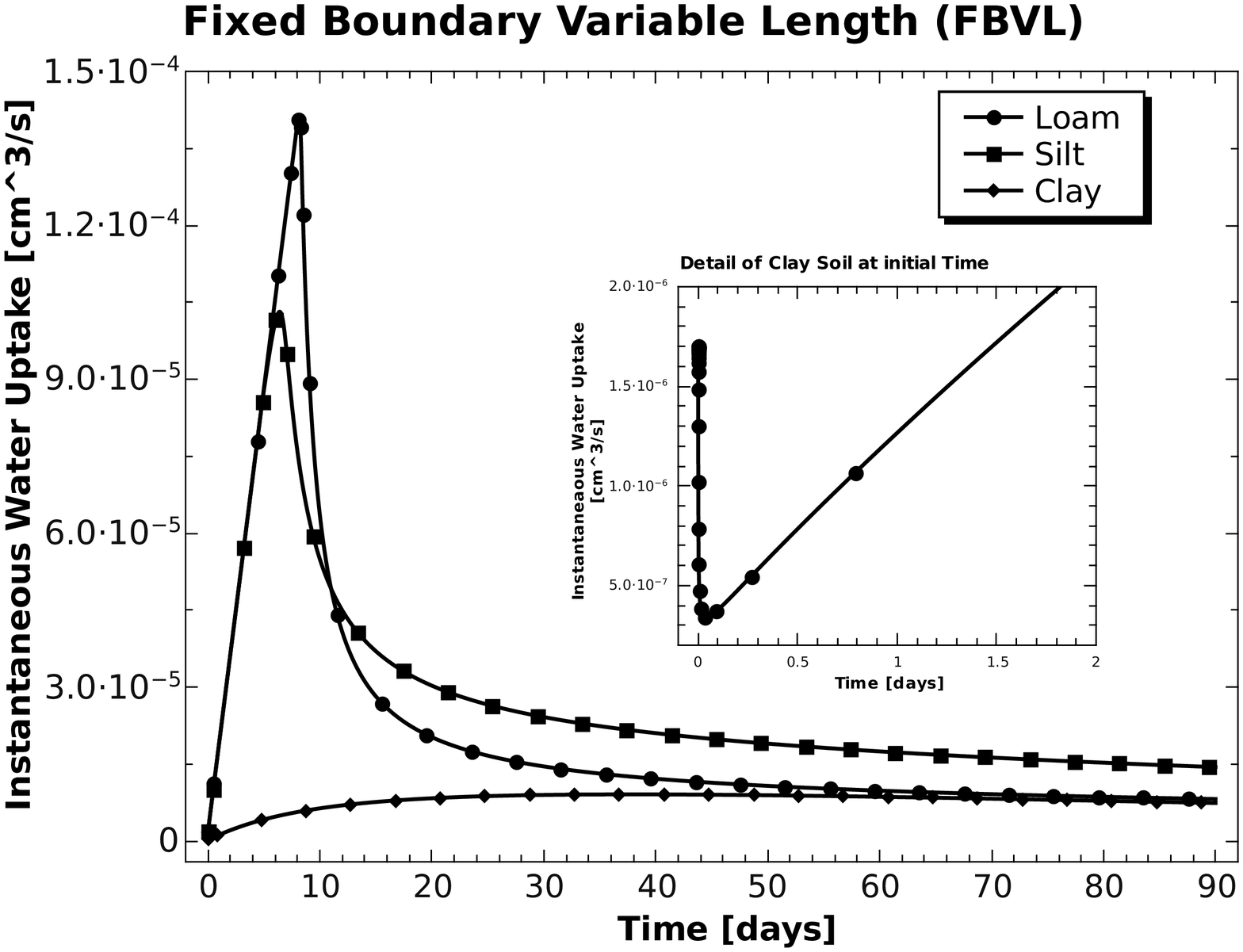}
      \\
      \includegraphics[scale=0.29]{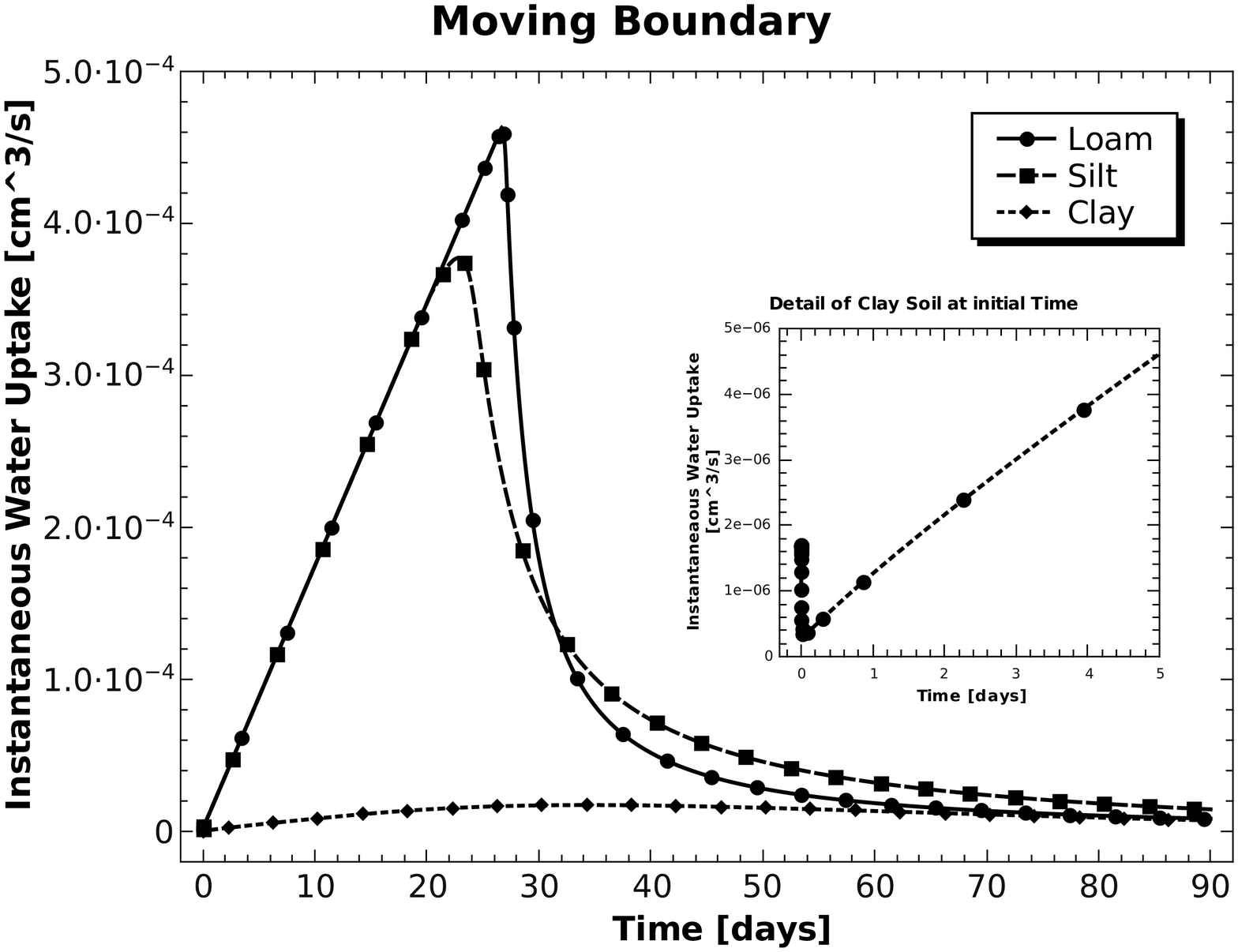}
    \end{tabular}
  \end{center}
  \caption{Instantaneous water uptake time evolution for the three models 
  to be  compared, from top to bottom fixed boundary without root growth, 
  fixed boundary with root growth and moving boundary by the root growth.}
  \label{fig:tinst}
\end{figure}

To compare the above simulations with those for a constant root length 
density in a fixed domain model a simulation with $V=0$ $(R(t) = R_0)$ was 
made. For the comparison to be useful the root length was taken to be the 
average root length $(l_0=\bar{l}=389.8cm)$. Using this value the domain 
is fixed and the moving boundary model then becomes the model proposed by  
Personne \cite{personne2003}. The influxes on the root surface estimated 
by the fixed boundary model are then integrated using  
(\ref{eq:tarziasimp}) with the fixed root length, to compute the 
cumulative uptake. This model will be referred to as Fixed Boundary Fixed 
Length (FBFL). Similar to published results on nutrient uptake (e.g. 
Claasen and Barber \cite{claasen1976}, Cushman \cite{cushman1979a}) the 
influxes obtained with the fixed boundary model can be integrated using 
equation (\ref{eq:tarziasimp}) with a variable root length, to compute the 
cumulative uptake. This model will be referred to as Fixed Boundary 
Variable Length (FBVL). Figure \ref{fig:tinst} shows the instantaneous 
root system water uptake against time (i.e. $G(\Psi(s_0,t)) \times l(t)$), 
for the three models. The time at which the instantaneous water uptake is 
maximum is called the Maximum Uptake time (MUT). The MUT is only present 
when the roots are growing. The straight line in the beginning is caused 
by the assumption of linear root length growth and the constant water 
uptake $(G(\Psi)=S_{max})$, on the models with root growth. In contrast 
when the root is taking up water in the water stressed (linear zone) of
the water uptake function the instantaneous water uptake as a function of 
time is non linear, in all models. We also observe that in the clay soil
for the models with root growth the water uptake initially decreases given 
the large decrease in soil water potential at the root surface. This
variation is shown in the inset, the later increase in water uptake is due 
to the root growth. The models with growing roots have a similar course in 
time, but the differences between the MUT-values are consistently showing 
that the MUT in the FBVL occurs earlier after the start of root growth. 
The instantaneous water uptake of the FBFL model and the FBVL models 
differs only by the multiplicative factor $l(t)/\bar{l}$.

\begin{figure}[p]
  \begin{center}
    \begin{tabular}{c}
      \includegraphics[scale=0.28]{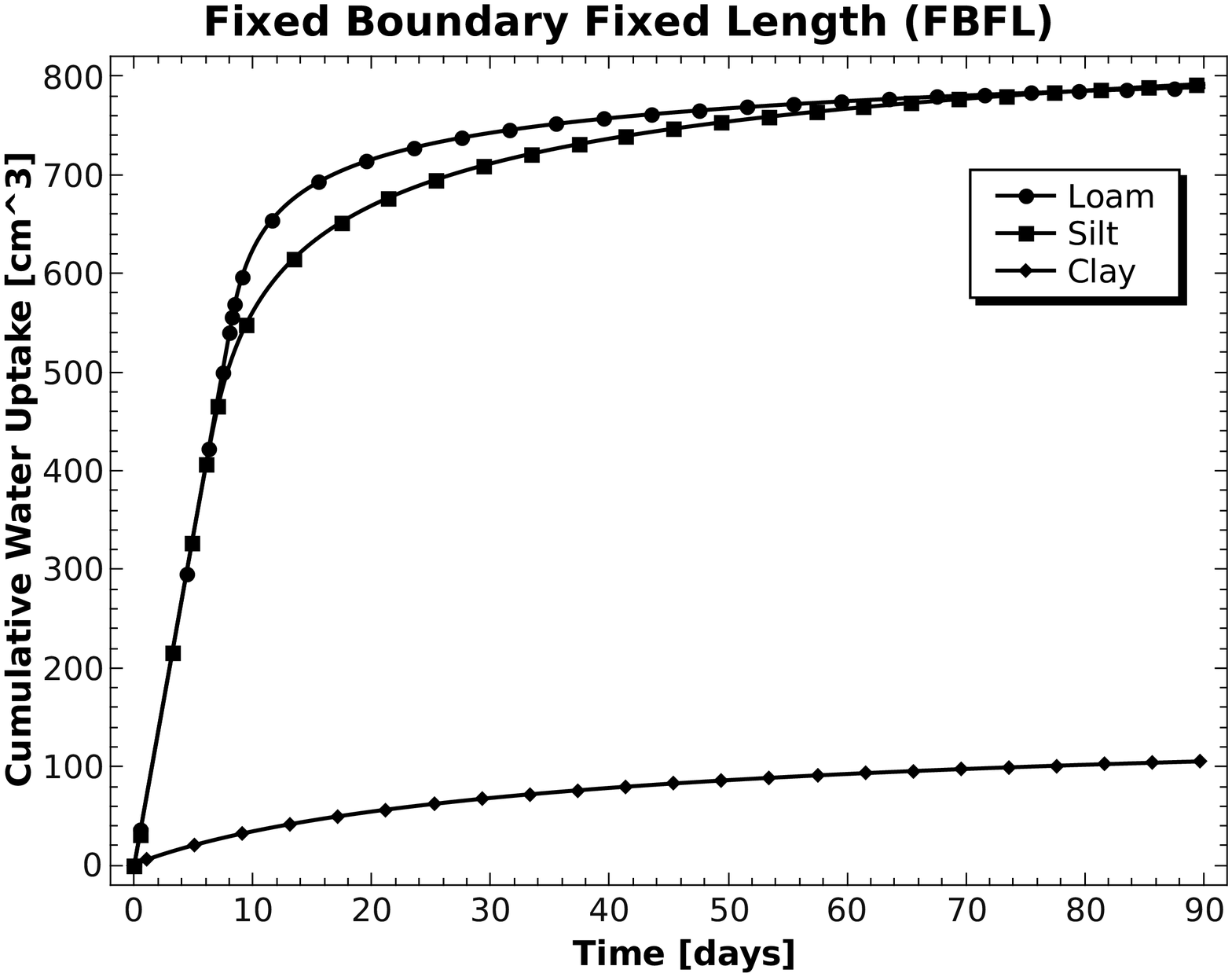}
      \\
      \includegraphics[scale=0.28]{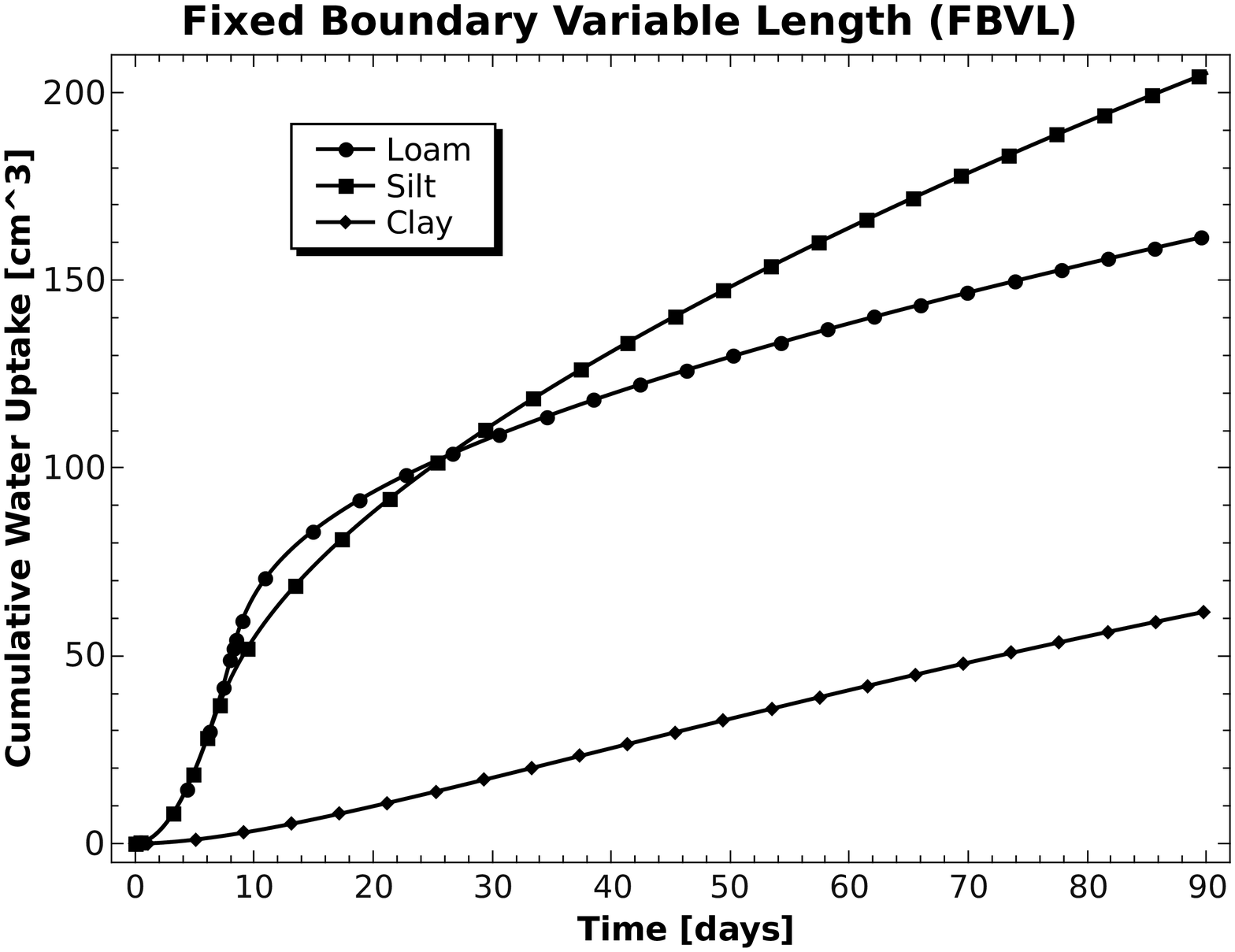}
      \\
      \includegraphics[scale=0.28]{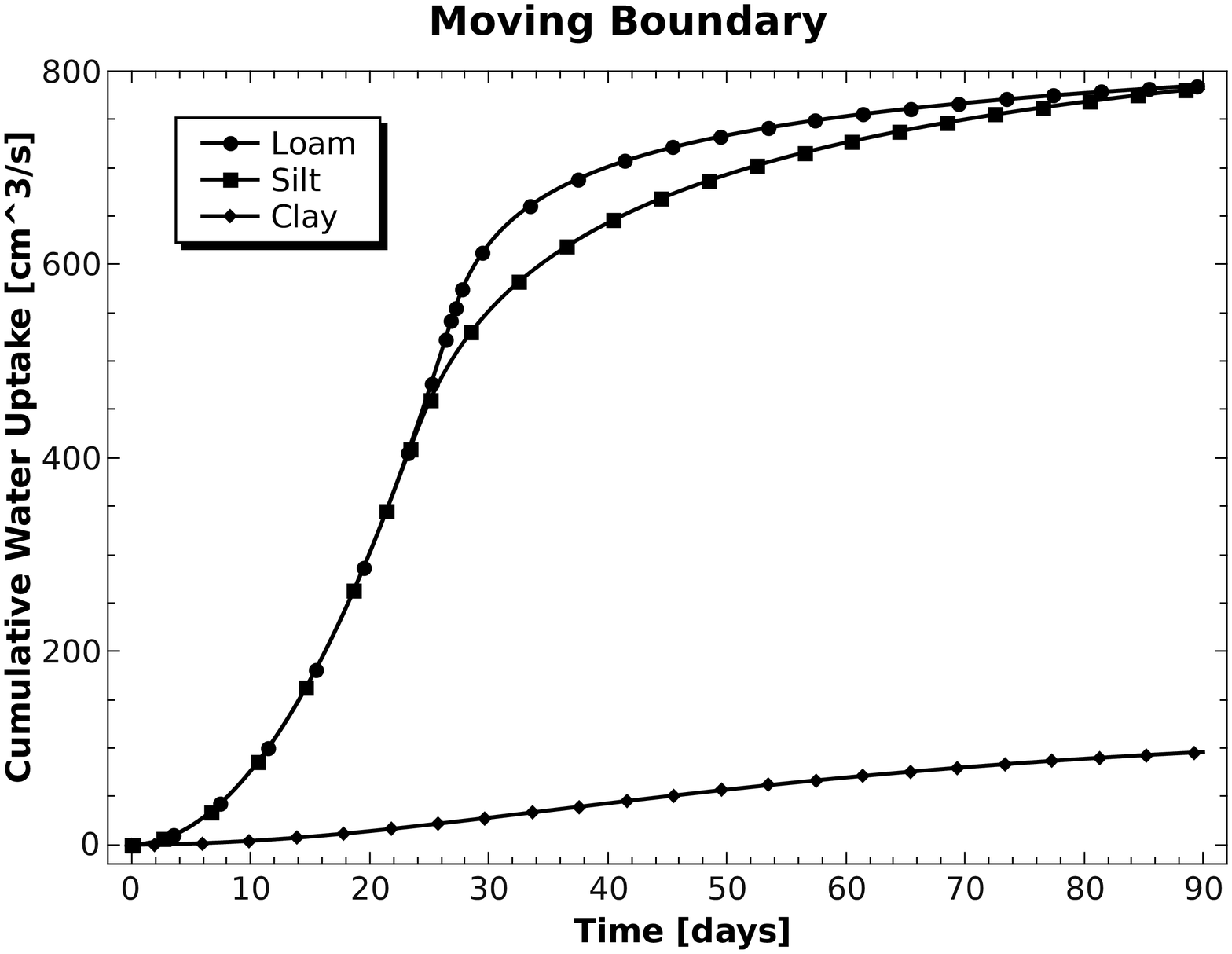}
    \end{tabular}
  \end{center}
  \caption{Cumulative water uptake time evolution  for the three models, 
  respectively fixed boundary without root growth, fixed boundary with 
  root growth and moving boundary with the root growth.}\label{fig:tacu}
\end{figure}

Figure \ref{fig:tacu} shows the cumulative water uptake as a function of 
time for the three models. The cumulative water uptake at 90 days 
$(U(90))$ is close to the amount of water initially available $(U_0)$ for 
the FBFL and the moving boundary models (see Table 
\ref{tab:parametros_iniciales}). For the FBVL model the cumulative water 
uptake is lower than that predicted by the other. This difference is due 
to the smaller value of the MUT. After 90 days $20\%$ of the initially 
available water in the clay soil is remaining, whereas for the silt and 
loam soil it is in the order of $10\%$. The clay soil does not trend to a 
constant value due to the less water uptake. It is expected that over a 
longer period the moisture content in the clay soil will asymptotically 
approach to a lower value, that might be on the proximity of the values found on the other soils.

To analyze the effect of parameter variation on  model output variables, 
several sensitivity diagrams were made using a local approach 
\cite{vangriensven2006}. The results are presented in Tables 
\ref{tab:sensibilidadl0}, \ref{tab:sensibilidads0},
\ref{tab:sensibilidadsmax} and \ref{tab:sensibilidadv} as a function of 
the multiplicative factor of each parameter for the loam soil. The more 
relevant parameters are $V$ and $S_{max}$, having both positive 
correlation on the water uptake at 30, 60 and 90 days and negative 
correlation with the MUT. As for the WUMUT, $S_{max}$ has a negative 
correlation and $V$ has a positive correlation. $s_0$ has an almost 
constant sensitivity and always positive correlation. Therefore the 
experimental measurement errors on those parameters would be more 
amplified on the output of the model. For the silt soil there is a very 
similar pattern than for the loam soil. For the clay soil there are some 
differences. $S_{max}$ is no longer a relevant parameter and $s_0$ is a 
relevant parameter with positive correlation to the accumulative water 
uptake. For the MUT and WUMUT $s_0$ has a negative correlation. The 
behavior of $V$ is left unchanged with respect to the loam soil. MUT shows 
almost no sensitivity to $S_{max}$ but the WUMUT shows that this parameter 
is very relevant with positive correlation. All the sensitivities are non 
linear with some exceptions for $s_0$.

\begin{table}[htp]
  \begin{center}
    \begin{tabular}{|c|c|c|c|}\hline
      Factor&$0.5$&$1.5$&$2.0$\\\hline\hline
      $U_{30}$&$-1.5\ 10^{-3}$&$1.5\ 10^{-3}$&$2.9\ 10^{-3}$\\
      $U_{60}$&$-1.6\ 10^{-4}$&$1.6\ 10^{-4}$&$3.2\ 10^{-4}$\\
      $U_{90}$&$-9.1\ 10^{-5}$&$8.7\ 10^{-5}$&$1.7\ 10^{-4}$\\
      MUT&$2.1\ 10^{-3}$&$-2.1\ 10^{-3}$&$-4.2\ 10^{-3}$\\
      WUMUT&$-7.8\ 10^{-5}$&$4.6\ 10^{-5}$&$1.1\ 10^{-4}$\\\hline
    \end{tabular}
  \end{center}
  \caption{Sensitivity of the moving boundary model to the variation of 
  the initial length $l_0$ for the loam soil. $U_{30}$, $U_{60}$, and 
  $U_{90}$ are the cumulative water uptake at 30, 60 and 90 days 
  respectively. MUT and WUMUT are the maximum uptake time and the 
  cumulative water uptake at maximum uptake time respectively. The 
  sensitivity is the change of the output divided by the value of the 
  output without variation of the parameter.}\label{tab:sensibilidadl0}
\end{table}

\begin{table}[htp]
  \begin{center}
    \begin{tabular}{|c|c|c|c|}\hline
      Factor&$0.5$&$1.5$&$2.0$\\\hline\hline
      $U_{30}$&$-1.5\ 10^{-2}$&$9.7\ 10^{-3}$&$1.7\ 10^{-2}$\\
      $U_{60}$&$-1.0\ 10^{-2}$&$6.6\ 10^{-3}$&$1.2\ 10^{-2}$\\
      $U_{90}$&$-7.7\ 10^{-3}$&$4.7\ 10^{-3}$&$8.1\ 10^{-3}$\\
      MUT&$-1.4\ 10^{-2}$&$9.0\ 10^{-3}$&$1.6\ 10^{-2}$\\
      WUMUT&$-2.8\ 10^{-2}$&$1.8\ 10^{-2}$&$3.2\ 10^{-2}$\\\hline
    \end{tabular}
  \end{center}
  \caption{Sensitivity of the moving boundary model to the variation of 
  the root radius $s_0$ for the loam soil. $U_{30}$, $U_{60}$, and 
  $U_{90}$ are the cumulative water uptake at 30, 60 and 90 days 
  respectively. MUT and WUMUT are the maximum uptake time and the 
  cumulative water uptake at maximum uptake time respectively. The 
  sensitivity is the change of the output divided by the value of the 
  output without variation of the parameter.}\label{tab:sensibilidads0}
\end{table}

\begin{table}[htp]
  \begin{center}
    \begin{tabular}{|c|c|c|c|}\hline
      Factor&$0.5$&$1.5$&$2.0$\\\hline\hline
      $U_{30}$&$-4.5\ 10^{-1}$&$6.4\ 10^{-2}$&$8.0\ 10^{-2}$\\
      $U_{60}$&$-1.3\ 10^{-2}$&$1.3\ 10^{-3}$&$1.5\ 10^{-3}$\\
      $U_{90}$&$-1.1\ 10^{-3}$&$1.0\ 10^{-3}$&$2.3\ 10^{-3}$\\
      MUT&$4.8\ 10^{-1}$&$-2.1\ 10^{-1}$&$-3.4\ 10^{-1}$\\
      WUMUT&$8.7\ 10^{-2}$&$-6.3\ 10^{-2}$&$-1.1\ 10^{-1}$\\\hline
    \end{tabular}
  \end{center}
  \caption{Sensitivity of the moving boundary model to the variation of 
  the maximum root water uptake per unit of root length $S_{max}$ for the 
  loam soil. $U_{30}$, $U_{60}$, and $U_{90}$ are the cumulative water 
  uptake at 30, 60 and 90 days respectively. MUT and WUMUT are the maximum 
  uptake time and the cumulative water uptake at maximum uptake time 
  respectively. The sensitivity is the change of the output divided by the 
  value of the output without variation of the parameter.}
  \label{tab:sensibilidadsmax}
\end{table}

\begin{table}[htp]
  \begin{center}
    \begin{tabular}{|c|c|c|c|}\hline
      Factor&$0.5$&$1.5$&$2.0$\\\hline\hline
      $U_{30}$&$-4.5\ 10^{-1}$&$1.0\ 10^{-1}$&$1.5\ 10^{-1}$\\
      $U_{60}$&$-5.9\ 10^{-2}$&$2.5\ 10^{-2}$&$3.9\ 10^{-2}$\\
      $U_{90}$&$-3.6\ 10^{-2}$&$1.6\ 10^{-2}$&$2.6\ 10^{-2}$\\
      MUT&$4.1\ 10^{-1}$&$-1.8\ 10^{-1}$&$-2.9\ 10^{-1}$\\
      WUMUT&$-7.3\ 10^{-3}$&$4.3\ 10^{-3}$&$7.4\ 10^{-3}$\\\hline
    \end{tabular}
  \end{center}
  \caption{Sensitivity of the moving boundary model to the variation of 
  the maximum root water uptake per unit of root length $S_{max}$ for the 
  loam soil. $U_{30}$, $U_{60}$, and $U_{90}$ are the cumulative water 
  uptake at 30, 60 and 90 days respectively. MUT and WUMUT are the maximum 
  uptake time and the cumulative water  uptake at maximum uptake time 
  respectively. The sensitivity is the change of the output divided by the 
  value of the output without variation of the parameter.}
  \label{tab:sensibilidadv}
\end{table}

\begin{figure}
  \begin{center}
    \begin{tabular}{c}
      \includegraphics[scale=0.3]{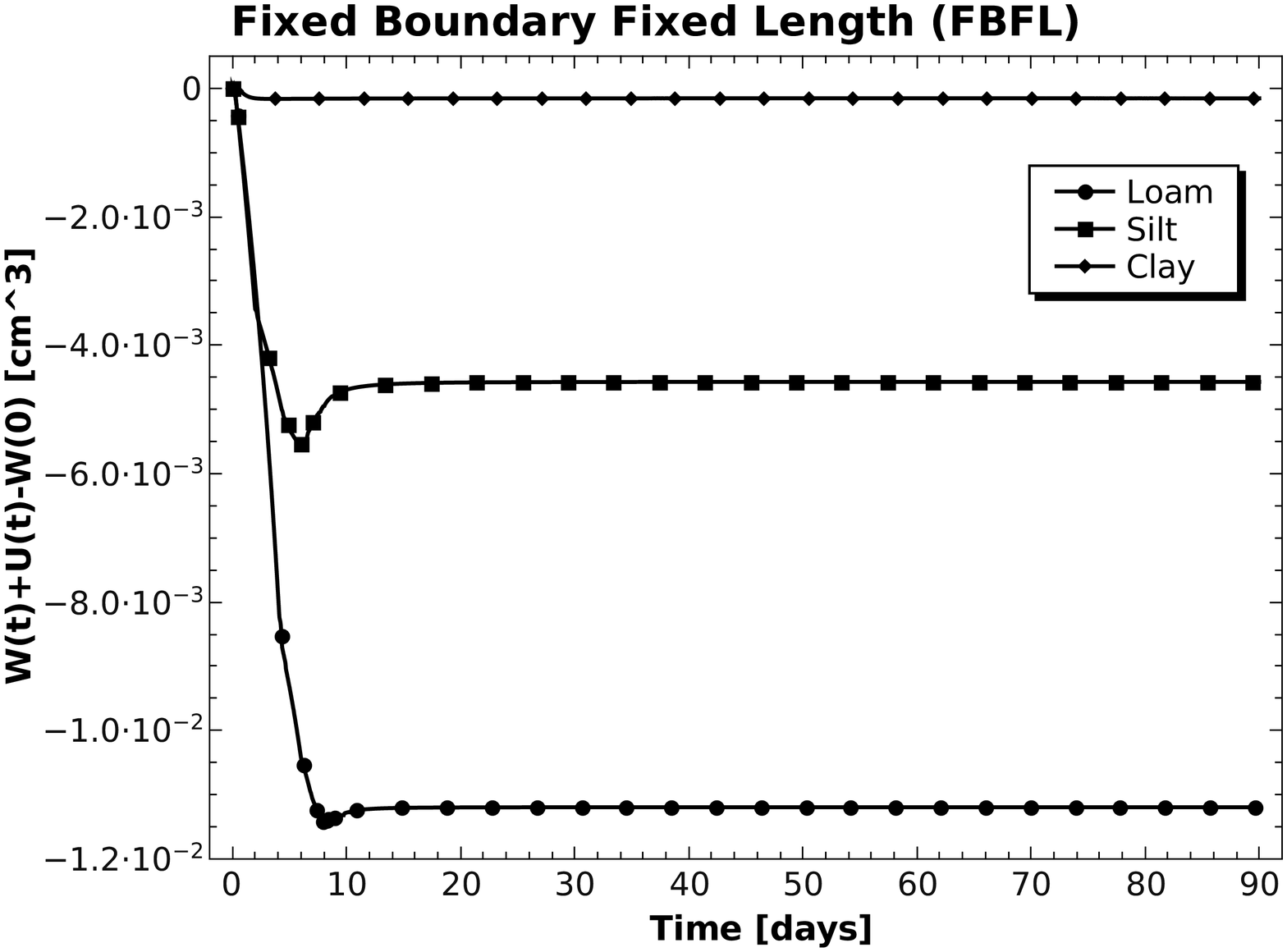}
      \\
      \includegraphics[scale=0.3]{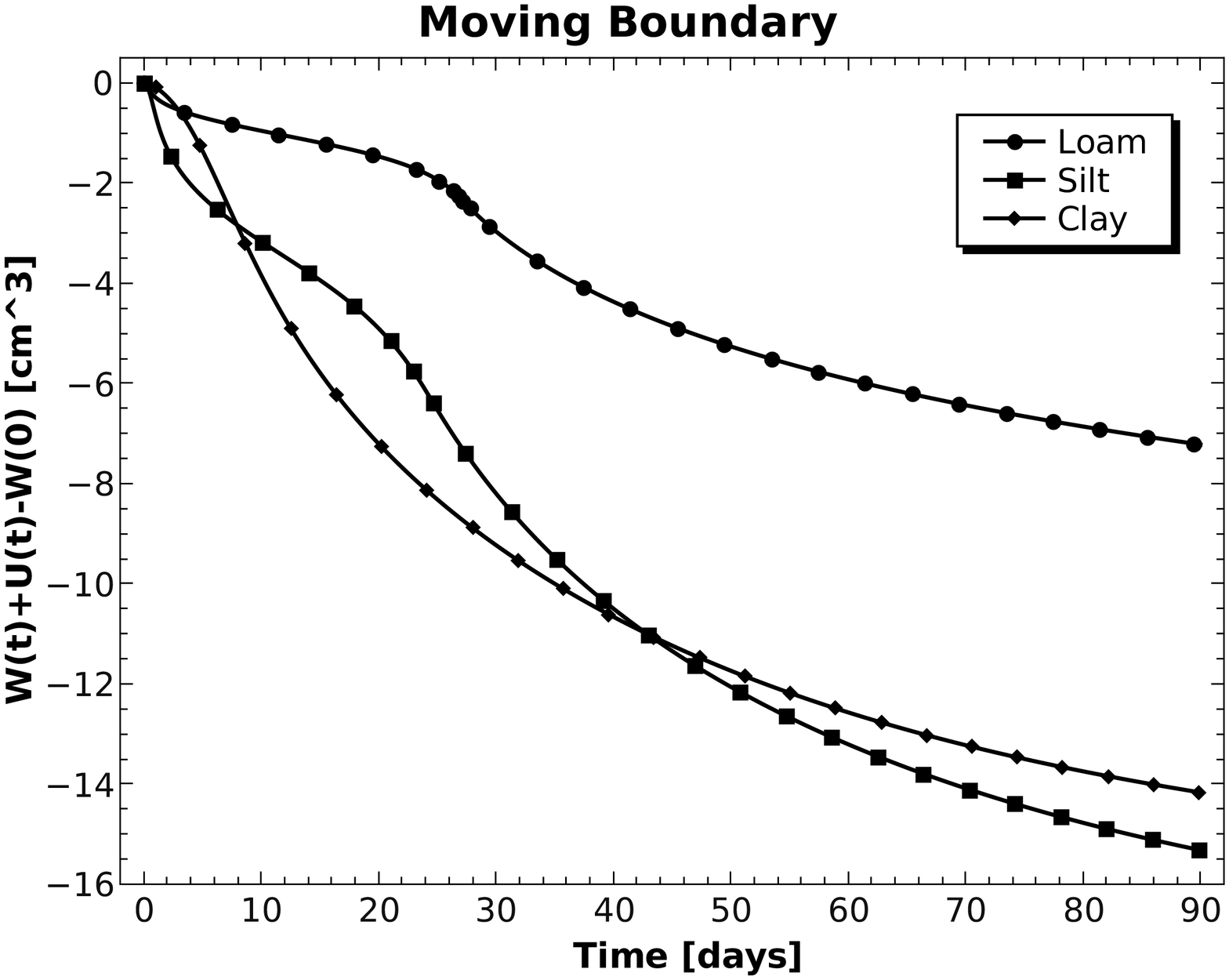}
    \end{tabular}
  \end{center}
  \caption{Water mass balance $(W(t)+U(t) -W(0))$ versus time $(t)$ for 
  the values shown in Table \ref{tab:tabla1} for the FBFL and the Moving 
  Boundary models. The total water is the cumulative water uptake plus the 
  water remaining in the soil, the initial water is the total water at the 
  start of the simulation.}\label{fig:balance}
\end{figure}

Mass conservation implies that the total water volume must be the same at 
the beginning and at an arbitrary time of the simulation. Since the water 
can only be in the soil or inside the root, therefore the total water 
volume $\left(T\left[cm^3\right]\right)$ at time $t$ would be
\begin{equation}
  T(t)= U(t)+W(t)
\end{equation}
with $U(t)$ and $W(t)$ defined as in equations (\ref{eq:tarziasimp}) and 
(\ref{eq:aguasuelo}) respectively. At the beginning of the simulation the 
cumulative water uptake is null, therefore the total water is the water in 
soil. With the above considerations the change of the cumulative water 
uptake plus the water remaining in the soil minus the initial water 
content $(W(t)+U(t) -W(0))$ as a function of time was calculated. For the 
FBVL model the mass balance cannot be done because to compute the water 
remaining in soil, the operation must be done with pressure head profiles 
as a function of time which has been calculated in fixed domain, but this 
result must be compared with the cumulative uptake by a growing root, 
which has been calculated integrating in a variable domain. The results of 
those calculations are shown in Figure \ref{fig:balance}. It shows that 
there is an effect of water mass loss for the moving boundary model that 
is not present on the fixed boundary model. There are two possible causes 
for this mass loss. The first cause is the assumption of total volume 
constant in the formulation of the model, reflected in the moving boundary 
formula $\left(R(t) = R_0 \sqrt{\frac{l_0}{l(t)}}\right)$, i.e., the 
volume that is kept constant is the soil plus root volume, being the 
volume occupied by the root is $V_R= \pi s_0^2 l(t)$. Then the soil volume 
is the total volume minus the root volume, therefore the root occupies a 
bigger fraction of the total volume as time increases (see Figure 
\ref{fig:evolucion}). The water content in the volume of soil removed by 
the root is not taken into account in the calculation of the water in the 
soil volume $(W(t))$. This means, in other words, that, for the model, the 
root is ``eating'' the soil, and the water which is in that portion of the 
soil. The second cause is the numerical errors. This case is similar for 
both models and does not contribute a large mass loss. An important remark 
is that the FEM is used to solve the water potential $(\Psi)$, not the 
water content $(\theta)$, on a discretized space with a finite precision, 
and cumulative errors could lead to a mass creation or loss depending of 
the parameters. Figure \ref{fig:balance} shows the soil volume loss effect 
plus the water mass loss due to the used numerical method in each 
integration step. In each soil for the moving boundary model the water 
loss is around $1\%$ of the total water. For the clay soil the total loss 
is around $15\%$ of the total water uptake, while for the other soils the 
loss is less than $2 \%$. The water loss for the fixed boundary model is 
negligible compared with the total water or the cumulative water uptake.

\section{Conclusions} 

An important remark is that all the results presented before are computed 
for a single plant, in a fixed soil volume with initial soil water content 
equal for all simulations, to compare the effect of the different soils on 
the water uptake. The parameters used represent traits of the plant. $V$ 
can be regarded as the response to nutrient and water uptake and is 
influenced by the plant genetics. $S_{max}$ is linked to atmospheric 
factors and to genetic variation. $s_0$ is the constant root radius. $l_0$ 
is a growing stage trait, changes in it only means a change in the 
development of the rooting system when the simulation begins.

There is a large difference in behaviour among the silt and loam soil on 
one hand and the clay soil on the other hand. Initial pressure head (Table 
\ref{tab:parametros_iniciales}) and pressure head profiles are different
(Figures \ref{fig:loam_profile}, \ref{fig:silt_profile} and 
\ref{fig:clay_profile}), and instantaneous water uptake is much smaller 
(Figures \ref{fig:sens_g_v} and \ref{fig:tinst}). This can be explained in 
terms of the water uptake function $(G(\Psi))$. For the clay soil the 
water uptake function is on the linear regime (zone (b) of Figure 
\ref{fig:tomafig}) and the water potential is sharply decreasing (drying 
the soil) at the root-soil interface. Once the time variation of the water 
potential at the interface is stabilized the root growth has the dominant 
effect on uptake and the curve is similar to the other curves, although it 
shows a non linear behavior.

From Figure \ref{fig:balance}, the mass balance is not exactly zero owing 
to the numerical method used. For the FBFL model the differences are due 
to numerical errors only and are very low compared with the water volume 
loss on the moving boundary model. The  mass variation is produced on the 
high instantaneous uptake zone, this is a known flaw of the finite 
elements method. The moving boundary model has a higher mass variation 
compared with the FBFL model by the accumulation of two factors, numerical 
errors as in FBFL model, and a soil volume loss effect. The soil volume 
loss effect is introduced by the assumption of constant total volume 
(roots grows at the expense of a decrease in the soil volume), which is 
reflected on the formula of the moving boundary (\ref{eq:modelo5}). For 
the moving boundary model the differences on the total values of water 
mass loss by a soil volume loss effect and numerical errors are due to the 
different variations of the soil water contents. The clay soil has low 
available water (see Table \ref{tab:parametros_iniciales}), but the total 
water mass loss is similar to the other soils, and it represents about $10 
\%$ of the initial total water. The assumptions that generate the soil 
volume loss effect should be revised when soils for which the initial 
potential is on the linear zone (i.e., low water availability) are 
studied. The water mass loss is consistent with the model assumptions, and 
the numerical induced errors are low compared to the water uptakes 
$(<15\%$ for clay soils, and $<2 \%$ for the other soils). There are two 
main actions to avoid the volume loss effect. The first one is to keep the 
soil volume constant and not the root-soil volume.  The second one is to 
evaluate the mass loss and put it back into the soil using a source 
function or a modification of the boundary conditions. Each procedure will 
have its advantages and disadvantages, but, since the water loss is very 
low compared with the water uptake for the studied cases, the revision is 
left for a specific work on low initial available water.

The obtained results shows a global consistency with the structure of 
soils studied, becoming a valuable tool to study the water uptake in a 
more complex situation (for example when the effect of a variable 
evapotranspiration on $S_{max}$ is considered).

Obviously, the results would change when plants not growing in pots but in 
the field (in this case equation (\ref{eq:modelo5}) is invalid) are 
considered. Here is necessary to develop a new rooting development 
function $R(t)$. One change in this model to be useful in a field 
situation is the use of a coordinate representing the depth and water 
transport by gravity. Here $R(t)$ should incorporate the root 
architecture. Moreover, since this is a first approach a simple water 
uptake function has been used. This function could be changed to a more 
complex function \cite{vrugt2001,li2006}. Moreover the effect of $V$ and 
$S_{max}$ could change substantially after the first 30 days in field 
conditions, since there will be circadian and climatic effects that will 
alters the root water uptake function $(G)$, and the evapotranspiration 
will be a parameter to be measured or calculated before the simulation.

The differences among models with and without root growth can bee seen in 
figures \ref{fig:tinst} and \ref{fig:tacu}. The dynamics of water uptake 
for the fixed boundary models are very different from those of a moving 
boundary model. The maximum instantaneous water uptake is achieved much 
faster for the fixed boundary model. Also a difference in the beginning 
and end of the simulation for the values of the instantaneous water uptake 
can be observed, while for the FBFL and the moving boundary model the 
instantaneous water uptake reaches values close to zero as time goes by, 
is not clear that the FBVL model has the same behavior. This difference is 
because the influx on root surface is computed for a single root in a 
fixed domain and the cumulative uptake by the growing root is calculated 
using these influxes. From a physical point of view, to compute the 
cumulative uptake influxes calculated in a variable domain should be used. 
Therefore the mass balance cannot be calculated for the FBVL. For those 
considerations of mass conservation the FBVL approach should be avoided on 
water uptake models.

\section{Acknowledgments}
The authors would like to acknowledge the valuable suggestions of 
Guillermo G. Weber from the University of Texas at Brownsville, and Sergio 
Preidikman from Universidad Nacional de C\'ordoba (UNC) for the 
implementation of the FEM. This work was supported in part by PIP N 
18/C362 from SECyT-UNRC  and PIP Nº 0534 from CONICET-UA, Argentina. The 
authors would like to acknowledge the valuable suggestions made by the 
anonymous referees to this work.

\bibliographystyle{elsarticle-num}
\bibliography{waterbiblio}

\appendix
\section{Deduction of simplified total water uptake}\label{app1}
The formula for water uptake given by \cite{reginato2002} is
\begin{eqnarray}
  U(t)& =&l_0\int_0^{t}G(\Psi(s_0,\tau))d\tau\nonumber\\&&+  \int_0^{t}
  \left(\int_\tau^{t}G(\Psi(s_0,\tau))d\tau\right) \dot l(\tau) d\tau,
\end{eqnarray}
the second integral on the right side can be reduced, by a parts 
integration, into
\begin{eqnarray}
  U(t)&=&l_0\int_0^{t}G(\Psi(s_0,\tau))d\tau\nonumber\\&&+\left.\left(
  \int_\tau^{t}G(\Psi(s_0,\tau))d\tau\right)l(\tau)\right|_0^t-\int_0^{t}
  \left(-G(\Psi(s_0,\tau))\right) l(\tau) d\tau,
  \\
  U(t)&=&l_0\int_0^{t}G(\Psi(s_0,\tau))d\tau\nonumber\\&&-\left(\int_0^{t}
  G(\Psi(s_0,\tau))d\tau\right)l(0)+\int_0^{t}G(\Psi(s_0,\tau))l(\tau) 
  d\tau
\end{eqnarray}
Therefore the final result is
\begin{eqnarray}
  U(t) &=& \int_0^t G(\Psi(s_0,\tau)) l(\tau) d\tau
\end{eqnarray}

\end{document}